\documentclass[10pt]{article}
\usepackage[T1]{fontenc}
\usepackage[portuges,english]{babel}
\usepackage{amssymb,latexsym,amsmath,color}
\usepackage[colorlinks,linkcolor=blue,urlcolor=blue,citecolor=black,
plainpages=false,pdfpagelabels,breaklinks]{hyperref}
\usepackage{palatino}
\usepackage{mathpazo}

\title{Potentiality and Contradiction in Quantum Mechanics}
\author{{Jonas R. Becker Arenhart}\thanks{Department of Philosophy, Federal University of Santa Catarina, Florianópolis, SC 88040-900, Brazil (jonas.becker2@gmail.com).} \and {D\'ecio Krause}\thanks{Department of Philosophy, Federal University of Santa Catarina, Florianópolis, SC 88040-900, Brazil (deciokrause@gmail.com)}}
\begin{document}
\maketitle

\newcommand{\ita}{\textit}

\begin{abstract}
Following J.-Y.Béziau in his pioneer work on non-standard interpretations of the traditional square of opposition, we have applied the abstract structure of the square to study the relation of opposition between states in superposition in orthodox quantum mechanics in \cite{are14}. Our conclusion was that such states are \ita{contraries} (\ita{i.e.} both can be false, but both cannot be true), contradicting previous analyzes that have led to different results, such as those claiming that those states represent \ita{contradictory} properties (\ita{i. e.} they must have opposite truth values). In this chapter we bring the issue once again into the center of the stage, but now discussing the metaphysical presuppositions which underlie each kind of analysis and which lead to each kind of result, discussing in particular the idea that superpositions represent potential contradictions. We shall argue that the analysis according to which states in superposition are contrary rather than contradictory is still more plausible.
\\
Key-words: contradiction; superposition; potentiality; contrariety; opposition.

\bigskip
\hfill{\textit{Dedicated to Jean-Yves B\'eziau for his 50th birthday.}}
\end{abstract}

\section{Introduction}

J.-Y.Béziau has advanced the thesis that the square of opposition is a general framework that may be profitably employed for conceptual analysis (\ita{e.g.} in Béziau \cite{bez12}). Almost any kind of opposition between propositions may be profitably studied by the conceptual machinery furnished by the square, so that the proper relationships between the propositions in question may be brought to light and further analyzed. In this sense, the traditional opposition between Aristotelian categorical propositions is but one of the many interpretations of the abstract structure of the square.

Bearing these multiple interpretations in sight, one of the possible uses of the square concerns application in the case of quantum superpositions. States in quantum mechanics such as the one describing the famous Schrödinger cat --- which is in a superposition between the states ``the cat is dead'' and ``the cat is alive'' --- present a challenge for our understanding which may be approached via the conceptual tools provided by the square. According to some interpretations, such states represent contradictory properties of a system (for one such interpretation see, for instance, da Costa and de Ronde \cite{cos13}). On the other hand, we have advanced the thesis that states such as ``the cat is dead'' and ``the cat is alive'' are contrary rather than contradictory (see Arenhart and Krause \cite{are14}, \cite{are14a}).

To keep the paper self-contained, we now recall the traditional definitions of the oppositions which shall be employed in this paper and that are used in the discussions of applications of the square:
\begin{description}
\item[Contradiction] Propositions $\alpha$ and $\beta$ are contradictory when both cannot be true and both cannot be false.
\item[Contrariety] Propositions $\alpha$ and $\beta$ are contrary when both cannot be true, but both can be false.
\item[Subcontrariety] Propositions $\alpha$ and $\beta$ are subcontraries when both can be true, but both cannot be false.
\item[Subaltern] Proposition $\alpha$ is subaltern to proposition $\beta$ if the truth of $\beta$ implies the truth of $\alpha$
\end{description}
Subalterns are not in any sense in a relation of opposition, but rather a kind of implication. Anyway, they are part of the traditional discussions of the square and so are usually included in the non-standard approaches to the square.

In this chapter we shall once again engage in the discussion by presenting some further differences between the two approaches to superposition, \ita{viz.}, the one that considers them as contradictions and the one that considers them as contraries. In the next section we present in outline both approaches, the one which considers quantum states of a superposition as contradictory and as contrary, so that our discussion can be self-contained. In section \ref{con} we discuss the relation between contradiction and potential properties. It is said that the analysis of superposition as contradictory holds when such a state is thought as a superposition of potential properties, so we analyze this claim. In section \ref{opp} we investigate whether a concept of potentiality can sit comfortably with a notion of contradiction. In view of the previous discussions, we conclude by defending the idea that contrariety is still a more adequate way to understand superpositions.

\section{Contradictions and contrariety in superpositions}\label{rev}

The proper understanding of superpositions is an open challenge typical of most of quantum mechanical conceptual innovations. Typically, to address the problem an interpretation of the theory is offered, and along with it, the hopes that the difficulties that are generated by superpositions in quantum mechanics get a proper explanation. As is well known, Dirac \cite[p.12]{dir11} claimed that superposition is not reducible to any classical notion, it is a \ita{sui generis} feature of quantum mechanics: it lies behind most of the applications of the theory, and represents somehow the essence of the novelties brought by quantum mechanics.

In a recent paper, da Costa and de Ronde \cite{cos13} proposed to deal with superpositions by adhering to a completely innovative approach to the superpositions and, as a result of such move, by adjusting the underlying logic with which we discuss such issues: according to their proposal, a superposition involves in general contradictory properties, and the underlying logic for the discussion of those issues is a paraconsistent logic. Indeed, it is usually said that paraconsistent logics deal with contradictions without allowing trivialization.\footnote{In a nutshell: in classical logic, in the presence of a contradiction any proposition whatever may be said derivable, and the resulting system is called \ita{trivial}. In paraconsistent systems, on the other hand, even if expressions \ita{formally representing} contradictions are derivable, not every proposition is also derivable, so that the system is not trivial. For the details see \cite{cos06}, and also the discussion in Béziau \cite{bez03}, \cite{bez14}.}

To take an example that illustrates the main thesis, consider a spin-$\frac{1}{2}$ system which is in the state $| \uparrow_z \rangle$. Now, when we change the direction and consider the $x$ axis, this state is in a superposition between $|\uparrow_x\rangle$ and $|\downarrow_x\rangle$. Each of these states corresponds to a projector operator $|\uparrow_x\rangle\langle\uparrow_x|$ and $|\downarrow_x\rangle\langle\downarrow_x|$, respectively, with each projector representing a propriety of the system (in this case, ``to have spin up in the $x$ direction'' and ``to have spin down in the $x$ direction'', respectively). Now, according to da Costa and de Ronde \cite[p.848]{cos13}, these properties ``which constitute the superposition and must be considered simultaneously are in general \ita{contradictory properties}''.

So, it seems that the idea is very simple. Two properties such as ``to have spin up'' and ``to have spin down'' (from now on, the context should make it clear that they are being taken in the same spatial direction), when their corresponding states are in a superposition, are said to be contradictory. Furthermore, they are not actual properties of the system, but rather possible or potential properties (more about this in the next section, see also de Ronde \cite{ron14}). Then, potentially, the system has contradictory properties (see da Costa and de Ronde \ita{loc. cit.}). According to them, this fact must be dealt with by a paraconsistent approach, that is, we must adopt a paraconsistent logic as the underlying logic (see \cite[sec.5]{cos13}).

Obviously, if two properties are to be thought of as contradictory, then one must make clear what is the meaning of contradiction. We attempted such an explanation in \cite{are14}. Indeed, if ``spin up'' and ``spin down'', or else (taking into account Schrödinger's cat) ``cat dead'' and ``cat alive'' are to be understood as contradictory, then this contradictoriness must be spelled out. To spell out this fact by using the traditional definition of contradiction of the square of opposition results in the fact that such attribution of properties must always have well defined truth values: recalling that contradictory propositions have opposed truth values, the system must have spin up \ita{and} not have spin down, or else have spin down \ita{and} not have spin up. However, this was obviously not the idea behind the claim that we must ``consider simultaneously'' those properties. It is simply impossible to take them simultaneously and still keep them as contradictories when ``contradiction'' is understood in the terms provided by the traditional definition of the square.

Then, an analysis through the square seems to provide for an obstacle for the consideration of contradictory properties in a superposition. What alternative do we have? One could, perhaps, insist in a paraconsistent approach along the following lines. Let us consider properties such as ``to have spin up'' and ``to have spin down'' as being somehow one the \ita{paraconsistent negation} of the other. Then, following the suggestion of the paraconsistent approach, we could explain how both can be taken as true of a system. Indeed, when the system is in a superposition we would have something of the form $\alpha \wedge \neg \alpha$, a true contradiction. However, this move would not do. First of all, an expression such as $\alpha \wedge \neg \alpha$ is not a contradiction in a paraconsistent logic, strictly speaking. As philosophical analyzes have made it clear (see for instance Béziau \cite{bez03} and \cite{bez14}), paraconsistent negations represent subcontrariety, that is, when $\neg$ is taken as paraconsistent negation, $\alpha$ and $\neg \alpha$ are subcontraries, not really contradictories. This move would at best amount to a change in terminology (see also Arenhart and Krause \cite{are14a}). Second, subcontrariety requires that both propositions involved can be true but both cannot be false. So, by adopting this view, one would be obliged to accept that in a superposition, at least one of the properties corresponding to the states in superposition always hold. This, however, is still weaker than the requirement that both always be the case. Furthermore, it is not clear whether this assumption does not violate some form of no-go theorems that prohibit such kind of property attribution in quantum mechanics. 

We have suggested an alternative route in \cite{are14}: to take common wisdom seriously and claim, following Dirac, that a superposition represents a new state of the system, one in which the system, as far as we know, does not have any of the properties involved. To explain how to account for superpositions according to this analysis, let us consider once again the case of the spin-$\frac{1}{2}$ system, as an electron in the $| \uparrow_z \rangle$ state. When we inquire about what happens in the $x$ direction, then the system is in a superposition between $|\uparrow_x\rangle$ and $|\downarrow_x\rangle$. Now, instead of allowing that the system has both associated properties $|\uparrow_x\rangle\langle\uparrow_x|$ and $|\downarrow_x\rangle\langle\downarrow_x|$, we say that the system does not have any of the properties. In this sense, it can be false that the system has ``spin up'' and it can be false that the system has ``spin down''. However, if it is the case that the system has one of the properties (\ita{e.g.} spin up), then it does not have the other (spin down). This situation describes precisely \ita{contrary} propositions. So, the case of such superpositions involves an opposition, but it is not contradiction, but rather contrariety.

Notice that this view does not rule out interpretations such as versions of the modal interpretation (see Lombardi and Dieks \cite{lom14}). It could be the case that the system, even if it is in a superposition, \ita{does} have one of the associated properties (modal interpretations, recall, break the eigenstate-eigenvalue link). However, as we mentioned, in this case, contrariety is still preserved, because only one of the properties is the case, while the other is not. In fact, modal interpretations seem to be incompatible with any kind of approach to superposition in which the states in superposition correspond to properties that must be taken simultaneously.

Now, even though this seems to be very plausible (to our minds, at least), the approach that considers that states in superposition are contrary does seem to take into account some assumptions about property instantiation that a paraconsistentist may deny. Indeed, de Ronde \cite{ron14} has approached the subject following these lines. According to this line of reasoning, to consider that states in superposition are contraries but not contradictories involves assuming an orthodox metaphysical view, which includes the assumption that quantum mechanics describes entities and how they bear properties. The paraconsistent approach, on the other hand, should be understood as taking properties as potentialities, following a completely different metaphysical approach. We shall explore the different metaphysical views underlying the paraconsistent approach in the next two sections.

\section{Potentiality and contradiction}\label{con}

As we have mentioned by the end of the previous section, perhaps two distinct kinds of metaphysical assumptions underlie the two analyzes proposed for the case of quantum superposition. As we have suggested, there may be a different set of metaphysical presuppositions making the job in each case, and these presuppositions may well be incompatible. So, in order to make things clearer, in this section we shall discuss a little more the suggestion that distinct modes of being underlie each kind of analysis, mainly by trying to bring to light the idea of potentiality that accompanies the paraconsistent approach.

As da Costa and de Ronde \cite{cos13} and de Ronde \cite{ron14} have suggested, systems described by states in a superposition are such that they have only potentially or possibly the properties associated with each state. More than that, those properties, as we have already quoted, must be taken simultaneously and are thought of as contradictory. Now, the main question is: what are possible properties or potencies and how they can be contradictory?

Let us begin by exploring the idea of a contradiction in the potential realm. Once we admit that reality is divided in two spheres, the actual and the potential (or possible), both equally real, we may have contradictions in both. Given that superpositions are existent only in the potential realm, we may concern ourselves only with this case. The first point we shall raise concerns terminology: possibility and potentiality are treated as synonymous, it seems, by the paraconsistent approach. However, ``possible'' here has two distinct senses. In the most straightforward sense, it is said of a proposition that it is possible, while in the intended sense we are discussing, it is said of a property that it is possible. It is obviously the second sense that is being used here: possibility regarding properties. However, in its traditional use, a possible property, also called a \ita{modal property}, is understood as a property that an object \ita{does not have, but could have}. It is a useful notion, for instance, in the metaphysical discussion concerning the numerical difference (or identity) of a statue and the piece of clay of which it is made: if modal properties are admitted as legitimate properties in this case, then the statue is different from the piece of clay. Clearly, the statute \ita{could not} be squashed and still be the same, while the piece of clay \ita{could} be squashed and still be the same (see Lowe \cite{low02}).

Now, if this analysis of a possible property is correct, then we are entitled to understand potential properties as metaphysicians have traditionally understood modal properties. However, if the properties corresponding to the states in a superposition are modal properties of the system, then there can be no contradiction between them. Indeed, once again consider the case of the electron in $| \uparrow_z \rangle$. When we are concerned with the $x$ direction, the system is in a superposition between $|\uparrow_x\rangle$ and $|\downarrow_x\rangle$, and both properties $|\uparrow_x\rangle\langle\uparrow_x|$ and $|\downarrow_x\rangle\langle\downarrow_x|$ are possible (modal) properties. However, if it is correct to associate possible properties with modal properties, then these are properties the system does not actually have. In this case, then, it is difficult to understand how they can be contradictory; both simply fail to be properties of the system, and as a bonus, our suggestion that they are merely contraries applies (recall section \ref{rev}).

So, modal properties, even though they may be related to potential properties, will not work. Let us leave modal properties behind for now and keep with the same line of inquiry, but now turning our attention to ``potential property''. It happens that one can advance a very similar argument. Traditionally, \ita{potential} is understood as being in straight opposition to \ita{actual} (more on this in the next section). However, once again, a potential property is one which is not actually possessed by the system (by definition). In a superposition, when we identify the properties corresponding to the states in the superposition as potential properties in this sense (\ita{i.e.} as opposed to actual), then, those are properties \ita{not} possessed by the system. Again, there is no contradiction in this case, but only mere contrariety.

Perhaps through these analyzes we are still considering the potential and the possible too closely related to actuality and actualization. What if we consider potentiality as a separated realm completely independent of any entity in which it exerts its actualization? Could this independence somehow help us in attributing some sense to the idea that a contradiction is real in this realm? It is this suggestion that we now investigate.

The first point of the suggestion seems to be that properties are potential all by themselves. In this sense, they are independent of their possible attribution to a system. That is, a property is not possible because it is the modal or potential property of any specific system, but rather it is an independent power, existent by itself. In this sense, powers like ``spin up'' and ``spin down'' are contradictory. But it is even more difficult to make sense of contradiction here.

As far as contradiction is defined by the square of opposition, or even in other contexts, it involves some kind of affirmation and some kind of negation. Also, it involves truth and falsehood. Both, broadly understood, are missing in the completely independent understanding of contradictory properties.

First of all, a property, taken by itself as a power (a real entity not actual), is not affirmed nor denied of anything. To take properties such as ``to have spin up in the $x$ direction'' and ``to have spin down in the $x$ direction'' by themselves does not affirm nor deny anything. To say ``to have spin up in the $x$ direction'' is not even a statement, it is analogous to speak ``green'' or ``red hair''. To speak of a contradiction, it seems, one must have complete statements, where properties or relations are attributed to something. That is, one must have something like ``spin up is measured in a given direction'', or ``Mary is red haired'', otherwise there will be no occasion for truth and falsehood, and consequently, no occasion for a contradiction. So, the realm of the potential must be also a realm of attribution of properties to something if contradiction is to enter in it. However, this idea of attribution of properties seems to run counter the idea of a merely potential realm. On the other hand, the idea of a contradiction seems to require that we speak about truth and falsehood.

Second, perhaps we can make clearer the idea of contradictory properties by analyzing the formal approach to contradictions advanced by da Costa and de Ronde. In \cite[p.855]{cos13} it is provided for a paraconsistent set theory $ZF_1$ in which superpositions are formalized. Let us consider a system $S$ which is in a superposition of states $s_1$ and $s_2$, both ``classically incompatible''. A predicate symbol $K(S, s_1)$ is introduced in the language of $ZF_1$ by da Costa and de Ronde to represent the predicate that ``$S$ has the superposition predicate associated with $s_1$''. The same reading holds with obvious adaptation for $K(S, s_2)$ and similarly for $\neg K(S, s_1)$ and $\neg K(S, s_2)$, where $\neg$ is a paraconsistent negation. Now, with the help of these predicate symbols the \ita{Postulate of Contradiction} is introduced: when $S$ is in a superposition of $s_1$ and $s_2$, we have $$K(S, s_1) \wedge \neg K(S, s_1) \wedge K(S, s_2) \wedge \neg K(S, s_2).$$ This postulate somehow represents the situation in a superposition.

Now, let us apply this postulate to our former example of an electron (our system $S$) in $|\uparrow_z\rangle$ which in the $x$ direction is in a superposition of the states $|\uparrow_x\rangle$ and $|\downarrow_x\rangle$ (our states corresponding to $s_1$ and $s_2$). The Postulate of Contradiction now reads: $$K(S, |\uparrow_x\rangle) \wedge \neg K(S, |\uparrow_x\rangle) \wedge K(S, |\downarrow_x\rangle) \wedge \neg K(S, |\downarrow_x\rangle).$$ But as we have remarked in \cite{are14}, the contradiction now comes from the postulation of $K(S, |\uparrow_x\rangle) \wedge \neg K(S, |\uparrow_x\rangle)$, not from any relation between $|\uparrow_x\rangle\langle\uparrow_x|$ and $|\downarrow_x\rangle\langle\downarrow_x|$. Anyway, let us concede this point and accept that a superposition is inconsistent because it involves things like $K(S, |\uparrow_x\rangle) \wedge \neg K(S, |\uparrow_x\rangle)$. There are some difficulties with this move.

To begin with, this attribution of contradiction requires that there is a system $S$ which both has ``spin up'' and does not have ``spin up''. This is clearly an actualist reading of property attribution, in the sense that the system has \ita{actually} the property and \ita{actually} does not have it. Indeed, this is a contradiction in the actual world. So, in this sense, the Postulate violates the requirement that the properties in a superposition be only potential in order to build a contradiction. This move clearly goes against the main motivation to consider the properties of a superposition as potential.

But what if we count such an attribution as merely potential? We could try to read $K(S, s_1)$ as ``the system $S$ potentially has the predicate associated with $s_1$''. Now, an ambiguity enters the stage in the case of negation. If this suggestion is correct, when we try to read $\neg K(S, s_1)$ we have two options. The first one reads ``the system $S$ does not have potentially the predicate associated with $s_1$'', understood as meaning that it is not potentially that the system has this property, it rather has it actually. The second reading is ``it is not the case that the system $S$ has potentially the predicate associated with $s_1$'', that is, it is false that $S$ is potentially $s_1$. Both readings are problematic.

The first reading is obviously troublesome when we consider the Postulate of Contradiction: it says both that $S$ has the predicates associated with $s_1$ and $s_2$ both potentially and actually. However, as the case of the electron illustrates, the system would have to be both ``spin up'' and ``spin down'' in the potential realm and in the actual realm, which, at least in the last case, is impossible. The second reading is also problematic for the approach. Indeed, since the motivation for approaching superpositions as contradictions was by considering properties in a superposition simultaneously, there seems to be no reason to consider them as holding simultaneously, even if potentially, and then deny that they hold simultaneously, even if potentially. That is, the introduction of a negation read as denying that the system has a given property potentially simply does not make sense if the idea was to represent a superposition as attributing both properties to the system (even if potentially). In this sense, the Postulate of Inconsistency does not seem to represent the intuition behind the paraconsistency approach to superpositions.

However, it may be the case that a paraconsistent set theory is just an inadequate formalism to capture the idea of a potential property, and a language involving an operator whose role is to represent potential property attribution could help us better in this task. Our next section shall investigate the prospects of this move.

\section{Potentiality and oppositions}\label{opp}

What if the difficulties presented above come from distinct senses of ``contradiction''? Perhaps the contradiction as represented in the traditional square is not the same as a contradiction for potential properties. In this section we shall discuss a little more the idea of a potential property by employing the square of opposition. Throughout this section, we shall introduce a special operator $\lozenge$ to represent potentiality. In this sense, $\lozenge p$ means that ``$p$ is potential''. Now, of course we must deal with $p$ as representing potential property attribution \ita{to something}, in general, to an already given system.

First of all, to establish the terminology, one could begin by distinguishing two distinct ways the operator may represent potential property attribution that could, both, represent the case of superposition. Let us suppose that $s_1$ and $s_2$ are in a superposition and let us concede in using $s_1$ and $s_2$ ambiguously both for the states as for the statements that the system is in the corresponding state. In a first reading of the situation, we could understand this situation as represented by the formula $\lozenge (s_1 \wedge s_2)$ or as $(\lozenge s_1 \wedge \lozenge s_2)$. We shall argue that only the second reading is a sensible reading of a superposition, and that it is difficult to understand any of both statements as contradictory.\footnote{We have our doubts about the possibility of representing a superposition this way, in particular, in reading the $+$ sign of a superposition as a conjunction, but we shall do that for the purposes of argumentation in this section (see also \cite{are14}).}

Consider first the second statement, $(\lozenge s_1 \wedge \lozenge s_2)$. If this is the intended meaning of the claim that a superposition involves potentiality, then clearly there is no contradiction in it. There is a simple analog in classical modal logic with the statement $(\lozenge p \wedge \lozenge \neg p)$. Notice that it is possible for some proposition to be the case and it is possible for it not to be the case, and this is different from the contradictory statement $(\lozenge p \wedge \neg \lozenge p)$. It is this last statement we would need to represent a contradiction, but this is clearly not the case in a superposition, that is, we do not have $(\lozenge s_1 \wedge \neg \lozenge s_1)$. Indeed, no one would claim that a state in a superposition stands for a potential property of the system and then try to represent such thing by including the information that it also \ita{does not stand for a potential property of the system}. That move, we believe, renders the project of introducing potential properties in the discussion senseless.

However, one may complain that this reading of potentiality is not the intended meaning at all. This meaning, it could be argued, behaves too close to classical potentiality, it does represent at best the relation of a potential property coming to actuality, and is not the real quantum potentiality (see de Ronde \cite{ron14}), which is pure potentiality independently of actuality. In this case, it is the first reading of $\lozenge$, as it appears in $\lozenge (s_1 \wedge s_2)$ that represents a superposition. If this is the case, then some further difficulties arise.

First of all, taken by itself, this does not represent a contradiction in the sense of the square. Indeed, one needs two statements in order to have such a contradiction. Perhaps the meaning of contradiction is different, as we have already suggested in the beginning of this section, in the sense that the paraconsistentist wishes that this formula be a logical contradiction, a formula that is always false (as defined in traditional logic textbooks). But what is the use in quantum mechanics of a formula that is always false? Furthermore, is it really the case that $\lozenge (s_1 \wedge s_2)$ is always false? Not really, in any standard version of normal modal logic this formula would represent a contingent statement, while in non-normal modal logic it would be always true in the non-normal worlds, for instance, due to the peculiar semantical understanding of $\lozenge$ in those worlds (see Hughes and Cresswell \cite{hug96}). So, to tackle this issue seriously one must provide the axioms or rules for the operator $\lozenge$, which we have been understanding only informally till now.

However, instead of advancing a formal analysis of the operator $\lozenge$, we shall now investigate different meanings it could have and see how well they fare in relation to the idea that $\lozenge (s_1 \wedge s_2)$ is a potential contradiction, or, a contradiction in the potential realm. To fix our ideas, let us keep with the case of the electron whose state is a superposition of $|\uparrow_x\rangle$ and $|\downarrow_x\rangle$. So, if the idea is that $\lozenge (s_1 \wedge s_2)$ represents a contradiction in the potential realm, in our example that means that potentially our system has both spin up and spin down. Obviously, no system has \ita{actually} both spin up and spin down.

Now, this raises some important questions. First of all, what is the relation of the operator $\lozenge$ with actuality? Let us introduce for simplicity an operator $@$ for actuality. One could think at first that actuality and potentiality are mutually incompatible: when something is potential it is not actual, and when it is actual, then it is not potential. In this sense, the statements $\lozenge p$ and $@ p$ are \ita{contradictory}, in the sense of the square. However, if this is a sensible reading, notice that even though $\lozenge$(spin up and spin down), we never have $@$(spin up and spin down). In this sense, the postulated contradiction in the potential realm never gets actualized, it does not work for actual entities. That leaves the contradiction in the potential realm unmotivated, and makes the reading of a superposition as ($\lozenge$ spin up and $\lozenge$ spin down) much more plausible, without the need for a contradiction, once again. That is, since the alleged contradiction is not doing any physical and any metaphysical work, we may plausibly leave it behind. Some people call it an application of Priest's razor, the metaphysical principle according to which we should not populate the world with contradictions beyond necessity (see Priest \cite{pri87} and Arenhart and Krause \cite{are14} for a related discussion of this principle in the same context).

For a second possibility, let us consider that $\lozenge$ and $@$ are not contradictory, but are somehow compatible, in the sense that potential properties may also be actual and vice versa. In this sense, $\lozenge p$ and $@ p$ represent rather \ita{subcontrary} statements, not contradictory statements. Indeed, according to this reading, any property must be either actual or potential (both cannot be false), but some properties can be both actual and potential (both can be true). Our main difficulty with this interpretation is once again the lack of motivation for introducing the claim that $\lozenge$( spin up and spin down) represents a contradiction and to prefer this reading of a superposition instead of ($\lozenge$ spin up and $\lozenge$ spin down), which is not contradictory in any sense. Because even though some simple properties may be understood as being both actual and potential, this is clearly not the case for quantum properties such as spin up and spin down: that is, even if one accepts that potentially the system can be both simultaneously, actually that never occurs. So, this reading would provide for a distinct understanding of the relation between $\lozenge$ and $@$ which saves the postulation of a contradiction in potentiality, but it still does not help motivating the postulation.

Obviously, our proposal is not that both $\lozenge p$ and $@ p$ could both be false, generating a kind of limbo between actuality and potentiality, but rather that \emph{if} one is going to concede that some kind of potentiality must be introduced to account for superpositions (notice the conditional), \emph{then} perhaps the best way to understand that potentiality in relation to superposition is by leaving $\lozenge (s_1 \wedge s_2)$ behind and sticking to $(\lozenge s_1 \wedge \lozenge s_2)$. In this sense, we can still grant that a superposition represents a state in which not both properties are actual, that is, it is not the case that $(@ s_1 \vee @ s_2)$ must be necessarily true. The case is left open whether one or the other obtains, but the issue depends on some interpretational details that we shall present very briefly in the next section.

\section{Conclusion: contrariety again}

We hope we have made it clear that even though a paraconsistent approach to quantum superpositions is viable and defensible, it is still hard to see it as well motivated by the theory. Furthermore, there are some difficulties related with the very idea that a paraconsistent logic deals with contradictions \ita{stricto sensu} as well as with the idea that superpositions are indeed contradictory (for the first difficulty, see again Béziau \cite{bez03}, \cite{bez14} and for the second see section \ref{rev} and \cite{are14}).

Now, to consider that a superposition is better understood as a contrariety still leaves some issues open. It is a position compatible with very weak requirements on quantum mechanics. Let us make the issue clearer. We take it as a rather reasonable assumption that whenever a quantum system is in an eigenstate, then it really does have the property associated with the corresponding eigenvalue. This is a \ita{Minimal Property Ascription Condition}, in a formulation taken from Muller and Saunders \cite[p.513]{mul08}:
\medskip
\begin{description}
\item[The minimal property attribution condition:] If a system is in an eigenstate of an operator with eigenvalue \textbf{v}, then the system has the qualitative property corresponding to such value of the observable.
\end{description}
\medskip
Notice that this puts a fairly weak condition for us to attribute properties to quantum systems: when in an eigenstate, we can surely say the system has the associated property. This is only half of the famous eigenstate-eigenvalue link. But what happens when the system is not in an eigenstate, when it is in a superposition? The condition is silent about that. One can complement the minimal condition in a variety of ways, for instance, by claiming that when not in an eigenstate the system does not have any of the properties associated with the superposition. This option is compatible with the claim that states in a superposition are contraries: both fail to be the case. Or instead of adopting this position, one can assume another interpretation, such as modal interpretations, and hold that even in a superposition one of the associated properties hold, even if not in an eigenstate (see Lombardi and Dieks \cite{lom14}). Following this second option, notice, the understanding of superpositions as contraries still hold: even when one of the properties in a superposition hold, the other must not be the case.

So, the idea that states in a superposition are contrary rather than contradictory are compatible with a variety of interpretational moves. It is compatible with assuming only the minimal property attribution condition. The paraconsistent approach, on the other hand, introduces a further interpretational postulate, a kind of converse for the minimal condition according to which every superposition corresponds to properties attributed simultaneously to the system (see \cite{are14}):
\medskip
\begin{description}
\item[Paraconsistent property attribution:] When in a superposition, the system does have the properties related to the vectors forming the superposition, and they are contradictory.
\end{description}
\medskip

This move, obviously, closes some alternative interpretation which are also plausible candidates. So, it seems, this is a further advantage of dealing with superpositions as contraries: one leaves open some important issues that are still hot issues of interpretation in quantum mechanics, while the paraconsistent property attribution seems to put \ita{a priori} constraints on the theory and its future developments.

\end{document}